\documentclass{article}
\usepackage[utf8]{inputenc}

\usepackage{amsmath}
\usepackage{amsthm}
\usepackage{amssymb}
\usepackage{amsfonts}

\usepackage{listings}
\usepackage[colorlinks]{hyperref}
\usepackage[nameinlink]{cleveref}

\title{Chain Rules for Hessian and Higher Derivatives Made Easy by Tensor Calculus}
\author{Maciej Skorski\\University of Luxembourg}
\date{November 2019}

\newtheorem{proposition}{Proposition}
\newtheorem{remark}{Remark}
\newtheorem{corollary}{Corollary}

\begin{document}

\maketitle

\begin{abstract}
Computing multivariate derivatives of matrix-like expressions in the compact, coordinate free fashion is very important for both theory and applied computations (e.g. optimization and machine learning).

The critical components of such computations are \emph{chain and product rules} for derivatives. Although they are taught early in simple scenarios, practical applications involve  high-dimensional arrays; in this context it is very hard to find easy accessible and compact explanation. 

This paper discusses how to relatively simply carry such derivations based on the (simplified as adapted in applied computer science) concept of tensors. Numerical examples in modern Python libraries are provided. This discussion simplifies and illustrates an earlier exposition by Manton (2012). 
\end{abstract}

\section{Introduction}

\paragraph{Chain Rules}
Consider functions $f(x)$ and $g(y)$ such that $f:\mathbb{R}^n\rightarrow \mathbb{R}$ and $g=(g^1,\ldots,g^n):\mathbb{R}^m\rightarrow \mathbb{R}^n$.
We will be interested in computing higher-order derivatives of the \emph{composed} function $f(g(y))$ in \emph{cartesian} coordinates. From now on let $D$ denote the (multivariable) derivative. In particular we are interested in the hessian
\begin{align*}
D^2 (f\circ g)  =  D^2_{y,y} f(g(y))
\end{align*}
For example for optimization problems we often want to know how the hessian of objective $f$ changes under reparametrization $g$.

The general form describing $D^2 (f\circ g)$ can be found with the use of product and chain rules. First we have
\begin{align}\label{eq:chain_rule_1}
D (f\circ g) = Df(g) \cdot Dg
\end{align}
Differentiating \Cref{eq:chain_rule_1} once more we get
\begin{align}\label{eq:chain_rule_2}
D^2 (f\circ g) = D(Df(g) \cdot Dg) = (D^2f(g)\cdot Dg) \cdot Dg + Df(g)\cdot D^2 g
\end{align}

\paragraph{Proper Understanding of Dot Product}
Note that dots in \Cref{eq:chain_rule_2} cannot be interpreted as matrix product. First, $D^2g$ is actually a 3-dimensional object (it stores second derivatives for each of the components of $g$). Second, the product $(D^2f(g)\cdot Dg\cdot Dg)$ cannot be simply evaluated as the matrix product because shapes (respectively $(n,n)$, $(n,m)$ and $(n,m)$) will not be compatible. 
A workaround might be to do derivatives \emph{elementwise}, that is computing
$\partial y^i \partial y^j$ for different combinations of $i,j$ in isolation. Not only this is tedious, but don't offer any simplifications or compact expressions~\cite{manton2012differential}. When applied to practical problems it makes computations orders of magnitude slower~\cite{laue2018computing}.

Fortunately, both issues can be easily fixed (and computations carried forward for higher dimensions) with little use of \emph{tensor products}. The critical concept is the idea of \emph{pairing} indices. 
The matrix multiplication 
\begin{align*}
    (A\cdot B)[i,j] = \sum_{k} A[i,k]\cdot B[k,j]
\end{align*}
can be seen as forming cross-products of elements such that indexing tuples match on chosen dimensions (here $(i,k)$ and $(k,j)$ match on the second dimension $A$ and first dimension of $B$) which we refer to as \emph{pairing}, and then aggregating over the common index which we refer to as \emph{reducing}. This operation can be naturally generalized to high-dimensional arrays (which we refer to as tensors), but with flexible choice of paired dimension as long as they are compatible (paired dimensions should have same values of possible indices/keys). 

We can also do dot products of more than two vectors at once and/or exchange the order of multiplication, provided that we do correct pairing.

In our case the pairing is quite intuitive, as we expect the output dimensions \emph{not} to contain $n$ (depend on $y$ and not on $x$).

\begin{proposition}\label{thm:main}
The dot products of the form
\begin{align*}
     (D^{\alpha}f(g))\cdot D^{\beta_{1}}g)\cdot D^{\beta_{2}g}\ldots
\end{align*}
which arise in  \Cref{eq:chain_rule_1} and higher-order generalizations should be understood as \emph{tensor dot product} where dimensions of the tensor $D^{\alpha}f$ which stores $n$ partial derivatives $\partial y$ are paired with dimensions of tensors $D^{\beta_j}g$ which run over $n$ components of $g$.
\end{proposition}
\begin{remark}[Utilizing symmetry]
Since mixed derivatives are symmetric (the fact often referred to as Schwarz's Theorem~\cite{minguzzi2015equality}), the order in which we pick dimensions of $D^{\alpha}f$ to pairing does not matter.
\end{remark}

\begin{corollary}[Chain Rule for Hessian]
Let $\mathbf{H}$ denote the Hessian matrix and $\mathbf{J}$ denote the Jacobian matrix. Then for $f$ and $g$ as defined above we have
\begin{align}
    \mathbf{H}(f\circ g) = (\mathbf{J}g)^T\cdot \mathbf{H}f(g)\cdot \mathbf{J}g + \sum_{k=1}^{n}\frac{\partial f}{\partial{y^k}}\cdot  \mathbf{H}g^{k}
\end{align}
where the dot product is understood as the standard matrix multiplication.
\end{corollary}

\paragraph{Known results / alternative approaches}

Formulas for high-dimensional derivatives, in the expanded form, were derived in the expanded form. Even then the problem remains non-trivial and errors happened even in respected textbooks e.g. see ~\cite{fraenkel1978formulae}.

The use of tensors for the same problem is studied in~\cite{manton2012differential} albeit more formalism is introduced and \Cref{eq:chain_rule_2} is modified rather than interpreted; also no numerical examples are provided - as we demonstrate care is needed as different implementations  follow different conventions.

The problem can be actually approached even easier when tensors are understood as in mathematical physics; the pairing and reductions become much easier when one distinguishes so called covariant (lower) and contravariant (upper) indices (the distinction comes from different behaviors under coordinate transforms). Essentially vector components are contra-variant whereas derivatives are covariant. In that language we write for example
the first derivative of $g$ as $\partial_j g^{i}$, the hessian of $f$ as $\partial_{j i}f$ (here we abbreviate the partial derivative with respect to the $i$-th part of the argument as  $\partial^i$). Then as a general rule only pair of covariant/contravariant indices can be paired. In our case we again pair components of $g$ with partial derivatives of $f$. The very usefull Einstein summation convention assumes the reduction (summation) is taken over paired index, so that we can write (equivalently to \Cref{thm:main})
\begin{align*}
    Df(g)\cdot Dg &= \partial_i f(g) \partial_j g^{i} \\
    D^2f(g)\cdot Dg\cdot Dg &= \partial_{j i}f(g) \partial_{k'}g^{i}\partial_{k''}g^{j}
\end{align*}
In our approach we use only the simplistic notion of tensors and dot products, which is widely adapted by applied computer science researchers. We hope that this discussion will be beneficial for those seeking for a compact explanation illustrated with working code.

\section{Preliminaries}

For our needs we think of tensors as multidimensional arrays
$a[i_1,\ldots,i_d]$, where the number of indices $d$ is called the dimension (or rank) of the tensor $a$.
This view, although over-simplistic, is broadly adapted in the applied computer science community~\cite{tensorflow2015-whitepaper,sympy} and suffices for our needs. Throught the paper we write components of vectors with upper indices $x=(x^1,\ldots,x^n)$.

\paragraph{Dot Products for Tensors}

The \emph{dot product} of tensors $a=a[i_1,\ldots,i_m]$ and $b=b[j_1,\ldots,j_n]$ over dimensions $p$ and $q$ is defined as 
\begin{align}
a\overset{i_p=j_q}{\cdot} b = \sum_{i_p=j_q}a[i_1,\ldots,i_p,\ldots,i_m]\cdot a[i_1,\ldots,i_q,\ldots,i_n]
\end{align}
Note that it is a tensor of dimension $m+n-2$, indexed by the choice of tuples $i_k$ where $k\in\{1,\ldots,m\}\setminus \{p\}$ and $j_k$ where $k\in\{1,\ldots,n\}\setminus\{q\}$.
We can think of it as a two-step operation
\begin{itemize}
    \item \emph{pairing products}: one forms products 
    $a[\mathbf{i}]\cdot b[\mathbf{j}]$ 
such that multiindices $\mathbf{i}$ and $\mathbf{j}$ match on two chosen dimensions
\item \emph{reducing by summation} over paired dimensions
\end{itemize}
While the first step forms a tensor of dimension $m+n$, the second operation reduces it by $2$. To appreciate the flexibility of that generalized product, note that in matrix multiplication 
we pair the last dimension of $A$ with the first dimension of $B$ (common $k$).

\paragraph{Derivatives}

If a tensor $a[i]$ (here $i$ can be a multiindex) depends on $u=(u^1,\ldots,u^n)$ (cartesian coordinates) then we define the derivative tensor
\begin{align}\label{eq:derivative_t}
 (D a)_{i,j} = \frac{\partial a[i]}{\partial u^j}    
\end{align}
This extends the tensor by one dimension. We follow the most common order convention: the existing dimensions of $a$ come first ~\cite{korelc2007acegen}. In particular when $g$ is a vector-valued function then $(Dg)_{i,j} = \frac{\partial g^i}{\partial x^j}$ is the (correctly shaped) Jacobian matrix.
It is important to stress that \emph{some implenentations may not follow this convention}, e.g. the popular Python package \texttt{SymPy}~\cite{sympy} puts the derivative dimension first, while the Mathematica package follows textbooks~\cite{korelc2007acegen}.

\section{Proof}
 
First we note that the product rule
\begin{align}
D(a\cdot b) = D a \cdot b + a\cdot Db
\end{align}
holds for tensors $a,b$ (as functions of some vector $u$ we differentiate over) and the tensor dot product. To prove it formally, fix two multi-indices $i$ and $j$ that are paired in the dot product. 
We can apply the standard product rule for (real-valued) functions $a[i](u)$ and $b[j](u)$ which gives us
\begin{align*}
    \frac{\partial (a[i](u)\cdot b[j](u))}{\partial u^k} = 
    \frac{\partial a[i](u)}{\partial u^k}\cdot b[j] + 
    a[i]\cdot \frac{\partial b[j](u)}{\partial u^k}
\end{align*}
Letting $u^k$ run over all components of $u$ we 
note that $\frac{\partial a[i](u)}{\partial u^k} = (Da)[i,k]$
and $\frac{\partial b[j](u)}{\partial u^k} = (Db)[j,k]$. 
Now take the sum over the matching pairs $i,j$ and exchange the order of differentiating and summation on the left.

Whenever we apply the chain rule to some expression of form $a(b(u))$ where $a$ and $b$ are vector-valued functions, the partial derivatives of $a$ with respect to the $k$-th variable appears together with the partial derivative of the $k$-th component of $b$. This is exactly the pairing in \Cref{thm:main}.

To see it formally and more generally, consider a tensor $a=a[i]$ (where $i$ might be a tuple/multiindex) where each component is a function of $z\in\mathbb{R}^n$. Suppose now that $z=b(u)$ is a function of $u\in\mathbb{R}^m$ (also represented as a tensor of dimension 1). The derivative of $a$ with respect to $u$ then becomes
\begin{align}
(D (a\circ v))_{i,j} = \frac{\partial (a[i](b(u)))}{\partial u^j} = \sum_{k=1}^{n}
\left.\frac{\partial a[i]}{\partial z^k}\right|_{z=b(u)}\cdot \frac{\partial b^{k}(u)}{\partial u^k}
\end{align}
This is equivalent to taking the dot product of the tensor 
$D a$ and $D b$ on the dimensions corresponding to components of $z$ and components of $b$ respectively.

We conclude \Cref{thm:main} by repeating the observations above multiple times.

\section{Numerical Examples}

\subsection{Rosenbrock function}

Recall that we write vector components with upper indices; we use brackets to denote powers. Take the Rosenbrock function (it is used as a performance benchmark in optimization)
\begin{align}\label{eq:Rosenbrock}
    f(x^1,x^2) = (1-x^1)^2 + 100\cdot ((x^1)^2-x^2)^2
\end{align}
Consider the transformation $y=g(x)$ where
\begin{align}\label{eq:Rosenbrock_reparam}
\begin{aligned}
g^1(x^1,x^2) &= x^1\\ 
g^2(x_1,x_2)&=(x^1)^2-x^2
\end{aligned}
\end{align}
\paragraph{Direct computations}
We have $f\circ g = (1-x^1)^2 + 100(y^2)^2$. Thus the Hessian matrix equals
\begin{align}
    \mathbf{H} (f\circ g) = \begin{pmatrix} 2 & 0 \\ 0 & 200 \end{pmatrix}
\end{align}
We compare this with the chain rule. Here we use \texttt{SymPy} library~\cite{sympy} to evaluate tensor dot products. Be careful about the non-standard convention for differentiating tensors!

We start with obtaining the Hessian directly
\begin{lstlisting}
import numpy as np
import sympy as sm
# symbols
x1,x2 = sm.Symbol('x1'),sm.Symbol('x2')
y1,y2 = sm.Symbol('y1'),sm.Symbol('y2')
# Rosenbrock map
f = (1-x1)**2+100*(x1**2-x2)**2
# coordinate transform
g1 = x1
g2 = x1**2-x2
# Rosenbrock after transform
f.subs([(x1,g1),(x2,g2)]).simplify()
\end{lstlisting}

\paragraph{Chain rule}

Now we can compare this with the chain rule
\begin{lstlisting}
### NOTE: SymPy puts derivative on first dim !!! ###

## tensors that appear in the chain rule
Jg=sm.derive_by_array([g1,g2],[x1,x2])
Jf=sm.derive_by_array(f,[x1,x2])
Hf=sm.derive_by_array(Jf,[x1,x2])
Hg = sm.derive_by_array(Jg,[x1,x2])

## first part of the chain rule: D^2f(g)*Dg*Dg
t1 = Hf.subs((x1,g1),(x2,g2))

# tensor product stores all (not paired) cross-pairs
# contraction pairs indices and sums up
t1 = sm.tensorcontraction(sm.tensorproduct(t1,Jg),[1,3])
t1 = sm.tensorcontraction(sm.tensorproduct(t1,Jg),[0,3])

## second part of the chain rule: Df(g)*D^2g
t2 = sm.tensorcontraction(sm.tensorproduct(Jf,Hg),[0,3])

## the output is correct: diagonal with entries 2 and 100
t1+t2
\end{lstlisting}

\bibliographystyle{plain}
\bibliography{references}
\end{document}